\documentclass[manuscript]{acmart}

\renewcommand\footnotetextcopyrightpermission[1]{}
\AtBeginDocument{%
  }

\usepackage{xspace}
\usepackage{color, colortbl}
\usepackage{array}
\usepackage{float}

\usepackage{graphicx}
\usepackage{multirow}

\begin{document}

\title[A Pandemic for the Good of Digital Literacy?]{A Pandemic for the Good of Digital Literacy? An Empirical Investigation of Newly Improved Digital Skills during COVID-19 Lockdowns}

\author{German Neubaum}
\email{german.neubaum@uni-due.de}
\orcid{0000-0002-7006-7089}
\affiliation{%
  \institution{Human-centered Computing and Cognitive Science, University of Duisburg-Essen}
  \country{Germany}
}
\author{Irene-Angelica Chounta}
\email{irene-angelica.chounta@uni-due.de}
\orcid{0000-0001-9159-0664}
\affiliation{%
  \institution{Human-centered Computing and Cognitive Science, University of Duisburg-Essen}
  \country{Germany}
}

\author{Eva Gredel}
\email{eva.gredel@uni-due.de}
\orcid{0000-0002-3689-9834}
\affiliation{%
  \institution{Institut für Germanistik, University of Duisburg-Essen}
  \country{Germany}
}
\author{David Wiesche}
\email{david.wiesche@uni-due.de}
\orcid{0000-0002-6086-1406}
\affiliation{%
  \institution{Institut für Sport- und Bewegungswissenschaften, University of Duisburg-Essen}
  \country{Germany}
}

\renewcommand{\shortauthors}{Neubaum et al.}



\definecolor{Gray}{gray}{0.9}
\newcommand{\mysubsec}[1]{\smallskip \emph{\textbf{#1.}}}
\newcolumntype{L}[1]{>{\raggedright\let\newline\\\arraybackslash\hspace{0pt}}m{#1}}

\begin{abstract}
 This research explores whether the rapid digital transformation due to COVID-19 managed to close or exacerbate the digital divide concerning users’ digital skills. We conducted a pre-registered survey with N = 1,143 German Internet users. Our findings suggest the latter: younger, male, and higher educated users were more likely to improve their digital skills than older, female, and less educated ones. According to their accounts, the pandemic helped Internet users improve their skills in communicating with others by using video conference software and reflecting critically upon information they found online. These improved digital skills exacerbated not only positive (e.g., feeling informed and safe) but also negative (e.g., feeling lonely) effects of digital media use during the pandemic. We discuss this research's theoretical and practical implications regarding the impact of challenges, such as technological disruption and health crises, on humans’ digital skills, capabilities, and future potential, focusing on the second-level digital divide.

\end{abstract}




\maketitle

\section{Introduction}
This paper aims to explore whether global challenges that lead to rapid digital transformation and enhanced use of technology (such as the COVID-19 pandemic) enable disadvantaged groups (e.g., older, female, and less educated individuals) within the population to catch up on their limited digital skills, being able to narrow the digital divide. To that end, we conducted a pre-registered survey among German Internet users to investigate a) which socio-demographic determinants predict newly improved digital skills during COVID-19, b) whether these skills were able to emphasize the psychological effects digital media use had on people’s lives during the pandemic and c) which digital skills individuals evaluate as necessary in light of the on-going digital transformation. 

In 2020, the respiratory tract disease COVID-19 spread throughout the world and most of daily communication occurred through digital channels. People started regularly participating in virtual meetings for work and for study, messengers and video conference tools helped them keep in touch with family members and friends, and online shopping services were used for daily purchases \cite{Juvonen_2021,Karl_2021,Koch_2020,lowenthal2020thinking,Nguyen_2021}. This accelerated digital transformation penetrating every area of people’s lives resulted in increased use of technology and, potentially leading to enhanced digital skills among the users.
According to the concept of the digital divide, there is an unequal distribution of access to digital technology across the globe (first-level digital divide), usage patterns and abilities (second-level digital divide), and effects of technology use (third-level digital divide; \cite{Scheerder_2017,van_Deursen_2018,Cosma_2020}. Therefore, once technology is available, people must have the skills to use these tools in a beneficial way. People’s digital skills are conceptualized as their ability to use technology in an effective, ethical, and confident way \cite{riina2022digcomp}. In fact, research has documented that some groups within a society are more likely to have more sophisticated digital skills than others. In the long run, this could lead to reinforcing spirals that those who have digital skills can participate and engage more and more in social and political life while those with less digital skills will isolate more and more \cite{Scheerder_2017,van_Deursen_2010}.

This need for acquiring digital skills has manifested itself in multiple digital literacy programs tested and evaluated in the field (e.g. \cite{Alt_2020,Lev_On_2020,Moore_2022,Nogueira_2021}. However, the circumstances in people's lives encourage or even force them to quickly acquire digital skills to adapt to a new situation. An obvious example is assuming new tasks as an employee on the job, for example, by needing to learn how to use new software or how to collaborate through a new social media platform \cite{Nikou_2022,moharrami2020role,Cetindamar_2024}. In many cases, people develop new or improve their digital skills through experience and self-learning.
A rather unique situation represented the global COVID-19 pandemic in 2020 until 2023. Early on, public health orders and physician recommendations suggested that individuals physically isolate themselves from each other to limit the spread of this serious respiratory tract disease \cite{Joseph_2022}. In the face of these recommendations, digital technologies allowing human communication despite being physically distant became key tools in people’s daily lives – for some for the first time, while others intensified their use \cite{Subramaniam_2021,Nagel_2020,Dwivedi_2020}. Therefore, people, for instance, started to engage more in videoconferences to exchange work-related information \cite{Karl_2021,mcclain2021internet}, to reconnect with friends and family members \cite{Juvonen_2021,Nguyen_2021,Scott_2022}, or even for emergency remote teaching at schools and universities \cite{Serhan_2020,lowenthal2020thinking}. Many individuals forcefully had to go from traditional to online shopping, also having to learn how to use online shopping services \cite{Koch_2020}. At the same time, individuals turned to online media to consume the dynamic state of knowledge on COVID-19, being challenged to differentiate between accurate and false information as well as reliable and less trustworthy sources \cite{gerosa2021mis,Gabarron_2021,Freiling_2021,Rocha_2021}. Given this increased intensity and modified way of using the Internet, 90\% of U.S. Americans, for instance, asserted that the Internet was essential or at least important for them personally during the pandemic \cite{mcclain2021internet}.

This rapid digital transformation of everyday communication required individuals to quickly acquire digital skills to participate in school, university, work, social, and political life. But did the digital transformation during COVID-19 lockdowns spread equally across the population?
In light of the evidence of the digital divide, it seems plausible to ask whether only those with high digital literacy benefited from the pandemic-induced digital transformation or whether this exceptional situation affected all parts of the population in each country, resulting in reducing inequalities in the distribution of digital literacy. Initial evidence documented digital inequalities referring to how individuals used the Internet during the pandemic: already advantaged individuals with socioeconomic and digital privileges (e.g., male, higher educated, with higher Internet skills) were more likely to engage in information and communication practices online during the pandemic than less advantaged individuals \cite{Nguyen_2021b,van_Deursen_2020}. Further corroborating these findings, U.S. Americans with higher educational degrees were more likely to use the Internet in new or different ways compared to those with lower educational degrees \cite{mcclain2021internet}. The living area seemed to also matter: U.S. Americans living in urban and suburban areas found the Internet more essential in their lives during the pandemic compared to those living in rural areas \cite{mcclain2021internet}. Similarly, data collected during the pandemic indicated that inequalities in the use of digital communication could be expanded in the context of the global pandemic as older people and those with fewer Internet skills were observed to reduce their digital communication during COVID-19 \cite{Nguyen_2020}. PEW research indicated that the vast majority (82\%) of individuals with lower tech readiness (i.e., having lower confidence in their abilities to use digital devices) expressed that the Internet
played an important role in their lives in the pandemic \cite{mcclain2021internet}.

Despite this enhanced use of technology throughout different socio-demographic groups, studies have not yet fully grasped the immediate and long-term psychological effects of the COVID-19 pandemic and the living conditions of individuals: Documented psychological effects have been deteriorations in mental health and psychological well-being in terms of increased anxiety, depression, posttraumatic stress symptoms, and feelings of loneliness \cite{Knox_2022,Weber_2022,Cooke_2020}. In fact, media and technology use has been corroborated to be a tool to escape some of these symptoms, to self-regulate, and to cope \cite{Eden_2020,Nabi_2022,Wulf_2022}. Still, these coping strategies were shown to have both, functional (e.g., recovering from stressful episodes) and dysfunctional (e.g., a distraction from problems) facets \cite{Wulf_2022} as well as positive (e.g., relieving pandemic-related stress) and negative (e.g., increased daytime tiredness or anxiety) effects \cite{Drazich_2022,Koban_2022,Sun_2023}. Further negative consequences have been perceiving a "Zoom fatigue" (as expressed by 40\% by U.S. Americans; \cite{mcclain2021internet}) in the sense of feeling worn out due to the number of video calls which, for instance, was found to be associated anxiety, depression, and stress among Turkish university students \cite{deniz2022zoom}. 
During and after the experience of enhanced digital media use during COVID-19, individuals might have formed or updated their opinions about which digital skills are necessary in today’s digitized society. The rapid COVID-19-induced digital transformation may have disclosed individuals’ lack of skills and personal challenges in terms of how to set up video conferences, judge online information and news sources, protect personal and private data, or impede encountering hate speech \cite{Bin_Naeem_2021,Buchholz_2020,Singh_2022}. 

We are, therefore, interested in technology users‘ views about which digital skills are needed in a digitally transforming society and who is in charge of fostering those skills. To address the above, we formulated the following research questions (RQs):\\
\textbf{RQ1: }Which socio-demographic factors are associated with improvements in digital skills during the pandemic? \\
\textbf{RQ2: }How did newly improved digital skills exacerbate or attenuate the psychological effects of the pandemic? \\
\textbf{RQ3: }Which are the digital skills of tomorrow from the public’s point of view?\\
\textbf{RQ4: }Who - in Internet users’ view - is responsible for fostering these digital skills?\\

This work aims to contribute towards understanding a) to what extent the digital transformation due to COVID-19 lockdowns closed or exacerbated the pre-existing divide among Internet users concerning their digital skills; b) whether newly improved digital skills throughout the pandemic helped individuals to buffer the negative psychological consequences of the unique living circumstances during COVID-19; and c) which digital skills users perceive as most needed in a digital society.

\section{Related Work}\label{sec:relatedwork}
\subsection{The multidimensionality of digital skills}
Since the emergence of digital communication, scholarship has dealt with the question of which type of knowledge and skills individuals need to use digital technologies in an effective and satisfactory way. This research covered a series of different concepts ranging from “computer literacy” \cite{horton1983information}, “digital competence” \cite{Janssen_2013}, “ICT literacy” \cite{Park_2020}, “information literacy” \cite{Johnston_2005}, “new media literacy” \cite{Ashley_2013}, “social media literacy” \cite{Cho_2022,Livingstone_2014,Schreurs_2020}, “algorithmic literacy” \cite{Oeldorf_Hirsch_2023} to "artificial intelligence (AI) literacy" \cite{Long_2020}. While these concepts slightly vary in their definition and focus, they all describe the abilities users need on a cognitive, technical, and emotional level to engage in digital technology use in a responsible way.
To provide a common and integrative understanding of digital skills and their dimensions, the European Commission coined the “digital competence framework for citizens” \cite{riina2022digcomp}. The corresponding definition of digital competence is the ability to use digital technologies in a confident, critical, and responsible form in different contexts such as learning, work, and civic participation. This ability requires the combination of knowledge (e.g., awareness of concepts and facts), skills (e.g., ability to perform tasks), and attitudes (e.g., particular mindsets/ways of thinking/feeling). The framework draws on five key competencies: 1) Information and data literacy: The ability to articulate information needs, to judge information and sources, and to organize information. 2) Communication and collaboration: The ability to participate in society and interact and/or work with others through technology. 3) Digital content creation: The ability to independently create and edit information and content. 4) Safety: The ability to protect one’s own and others’ privacy and data, as well as physical and psychological health. 5) Problem-solving: The ability to detect and resolve problems in digital environments and to keep up with technological evolution.

\subsection{The unequal distribution of digital skills}
In the face of an increasingly dynamic evolution of technology, it is a key challenge to constantly adapt the digital competence framework to grasp the necessary skills and to ensure that these skills are spread equally among technology users. The latter challenge is closely tied to the notion of the “digital divide,” describing a widening gap between those who have access to technology (first-level digital divide), skills and literacy when using technology (second-level digital divide), effective ways to engage and benefit from technology use (third-level digital divide) and those who have not \cite{Cosma_2020}.
While first-level digital divide remains a global problem due to inequalities regarding access to devices, hardware, software, and subscriptions and means to maintain those \cite{Cosma_2020}, second- and third-level divides comprise more complex debates about how the inequalities look like, what their roots are, and with which interventions they can be resolved \cite{van_Deursen_2015,Scheerder_2017}. Most studies focused on socio-demographic determinants of the second-level digital divide in terms of who uses the Internet the most versus the least \cite{Scheerder_2017}. When it comes to the distribution of different forms of digital skills, studies showed a recurring pattern: younger, male and higher educated users have more sophisticated digital skills than older, female, and lower educated ones when using digital technologies (e.g., \cite{Correa_2015,Mart_nez_Cantos_2023,Oeldorf_Hirsch_2023,_zsoy_2020,van_Deursen_2010}. This rather consistent pattern of a divide in digital skills can foster societal inequalities by impeding that female, elderly and lower educated citizens participate in social and political processes, sometimes even leading to their social isolation and lack of representation. Consequently, the question arises of how this second- and third-level digital divide can be closed.


\section{Method}\label{method}
To answer our research questions, we conducted a pre-registered survey among German Internet users which was approved by the local Ethics Committee on 30 June 2023 (protocol no. 2306PBNG2372). The survey was carried out online over a time period of five days (August 2023). To facilitate the survey and to recruit participants, we used an online commercial panel (Bilendi GmbH\footnote{\url{https://www.bilendi.de/}}) that invited adult German panelists per predefined quotas (see \ref{sec:participants}) representative of German Internet users in terms gender and age. On average, completing the survey took participants 10,52 minutes. After completion, survey participants received monetary compensation (1,40 EUR). The data set and pre-registration of this study can be accessed online\footnote{\url{https://osf.io/su7fb/?view_only=3a362cbf83bc4ed28c3e142975c84788}}. In the following, we present the survey instrument that was used (section \ref{sec:instrument}) and information regarding the participants (section \ref{sec:participants}).

 \subsection{Survey Instrument}\label{sec:instrument}
The survey instrument consisted of three thematic sections: a) improved digital skills during the COVID-19 pandemic; b) digital media’s intervening influence on pandemic-induced psychological effects; c) future digital skills. Next, we describe the items and scales for each section. 

\textbf{Improved digital skills during the COVID-19 pandemic.} The measurement of newly improved digital skills was adapted from previously used instruments related to the EU digital competence framework asking participants to estimate their abilities to conduct a series of different online activities (such as \cite{rubach2019skala,rubach2023systematic,tzafilkou2022development, riina2022digcomp}). Therefore, we covered the dimensions "information and data literacy" and "safety" by implementing the factors "information seeking" and "technical knowledge," while we covered "communication and collaboration" through the dimension "communication." We extended the list by including educational activities and online shopping as these activities were found to be prevalent in digital media use during the COVID-19 pandemic \cite{Serhan_2020,lowenthal2020thinking,Koch_2020,gerosa2021mis,Gabarron_2021,Freiling_2021,Rocha_2021}. After briefing participants that the pandemic increased people’s use of digital technology, we asked "which of the following digital skills did you newly develop or improve throughout the pandemic, that is, during the lockdowns?” offering a five-point scale with 1 = “no change compared to before the pandemic”, 2 = “minimally improved”, 3 = “somewhat improved”, 4 = “improved”, and 5 = “significantly improved compared to before the pandemic.” Thus, we developed 32 items divided in five different dimensions. These were information seeking (six items such as “to critically reflect upon information I read online”; Cronbach’s $\alpha = .94$), communication (nine items such as “to communicate with others by using video conference software (such as Zoom or Teams)”; Cronbach’s $\alpha = .93$), commercial use (four items such as “to manage my finances using online banking”; Cronbach’s $\alpha = .93$), personal education (five items such as “to find appropriate formats (e.g., online courses) to educate myself”; Cronbach’s $\alpha = .90$), and technical knowledge (eight items such as “to protect my data online (e.g., using privacy settings)” or “programming”; Cronbach’s $\alpha = .95$). While "information seeking" referred to skills on how to deal with online information (e.g., critically reflecting upon the credibility of information and sources), “personal education” covered abilities to find sources of information (e.g., either through Wikipedia or online courses). 
 
\textbf{Digital media’s intervening influence on pandemic-induced psychological effects. }To assess to what extent digital media were an intervening force in how COVID-19 affected the individuals, we asked, “To what extent did digital media contribute to your feeling as follows during the pandemic?” Although RQ2 asked whether newly developed digital skills modified the psychological effects of the pandemic, this specific survey question did not relate to newly developed skills by purpose as we did not want to overstrain participants. It would have been rather difficult for participants to trace back whether some effects are attributable to new skills or to the mere use of digital technology in general. Thus, we decided to address RQ2 by combining an easier question for participants on perceived effects of digital media use with a correlation analysis (as pre-registered) with the variable of newly developed digital skills in order to disburden participants. Using 16 items, we measured psychological effects drawing on self-determination theory \cite{deci2012self} and the fulfillment of basic psychological needs \cite{kermavnar2024assessing} covering a diversity of psychological states such as feeling “competent,” “informed,” “safe,” “satisfied,” “connected with others,” “lonely,” or “inspired” on a five-point scale ranging from 1 = do not agree at all to 5 = fully agree. To reduce the number of items to factors, we computed an exploratory factor analysis (principal axis analysis with varimax rotation) that based on the empirical eigenvalues and a parallel analysis suggested a two-factor solution. We labeled these two factors as “positive effects” (eleven items; Cronbach’s $\alpha = .92$) and “negative effects” (five items; Cronbach’s $\alpha = .86$).
 
\textbf{Future digital skills.} Prompting participants to think about the future, we asked “which digital skills will the people of tomorrow need?” offering a five-point scale from 1 = absolutely not necessary and 5 = very necessary. 16 items derived from the digital competencies framework \cite{riina2022digcomp} covered aspects such as “to estimate the consequences of one’s actions on the Internet,” “to virtually communicate with others in a polite and an appropriate way,” “to handle artificial intelligence such as ChatGPT,” or “to distinguish serious from unserious sources on the Internet.”

\textbf{Responsibility to foster digital skills.}
We measured participants’ evaluation of the responsibility to foster digital skills on two levels: 1) status quo: “Please sort in a decreasing order who – currently – is in charge to foster people’s digital skills?” and 2) ideal status: “Please sort in a decreasing order who should be in charge to foster people’s digital skills”. We covered the following: “individual user,” “teacher / school,” “parents,” “news media,” “organizations (e.g., media regulatory authority,” “politics through regulation and curricula,” “local authorities (e.g., adult education center),” and “platforms / Internet services”. Higher numbers represent a higher responsibility.

\subsection{Participants}\label{sec:participants}

We collected data from 2,016 German Internet users in August 2023. Among those, n = 873 were not considered in the analyses as n = 38 rejected giving their informed consent, n = 359 terminated their survey before they reached the last page, while n = 476 failed the quality test. We implemented the latter by asking participants to check a certain category of the scale. A final sample of N = 1,143 individuals (562 identified as female, 575 identified as male, 6 identified as “other”) completed the questionnaire. The age of the sample ranged from 18 to 87 years (M = 46.69, SD = 15.81). Given that the online commercial panel invited participants based on pre-defined quota, our final sample resembles the distribution of age and gender of German Internet users \cite{GIK2022}: 18-29 years (m: 9.8\%, f: 9.7\%), 30-39 (m: 9.0\%, f: 8.6\%), 40-49 (m: 7.9\%, f: 8.7\%), 50-59 (m: 11.0\%, f: 11.02\%), 60-89 (m: 12.6\%, f: 11.1\%). In terms of educational level, 24.0\% had a low level of education (no or low/middle school qualification), 44.2\% had a moderate level of education (high school qualification or completed apprenticeship), and 30.5\% had a high level of education (college degree or doctoral qualification). 60.5\% lived in rural areas (countryside or small towns), while 39.5\% lived in urban areas (i.e., large cities). In terms of socio-demographic information, we also measured income, marital status, and living situation (alone vs. with others).
The digital devices most participants used daily were smartphones (88.6\%), notebooks (40.3\%), and desktop computers (33.2\%). Participants’ most frequent activities while using the Internet (measured on a scale from 1 = never to 6 = daily) were reading/writing emails (M = 5.38, SD = 0.93), using search engines as Google (M = 5.36, SD = 0.92), reading newsletters (M = 4.37, SD = 1.65), using online maps such as GoogleMaps (M = 4.11, SD = 1.31), using online encyclopedias such as Wikipedia (M = 3.94, SD = 1.42) and online news sites (M = 3.91, SD = 1.84). The social media technologies participants used most frequently (measured on a scale from 1 = never to 6 = daily) were WhatsApp (M = 5.25, SD = 1.53), YouTube (M = 4.23, SD = 1.68), Facebook (M = 3.73, SD = 2.19), and Instagram (M = 3.43, SD = 2.25).

\section{Results}

\subsection{Prevalence of improved digital skills due to the pandemic}
Drawing on the five focal factors, we see that the area Internet users’ most likely improved skills throughout the pandemic, was information seeking (M = 2.48, SD = 1.23), followed by commercial use (M = 2.24, SD = 1.26), communication (M = 2.13, SD = 1.08), personal education (M = 2.12, SD = 1.10), and technical knowledge (M = 1.92, SD = 1.04). When inspecting the individual improved skills (by collapsing the two levels of “improved” and “significantly improved”), it becomes clear that besides improving their skills “to communicate with others by using video conference software (such as Zoom or Teams)” (34.2\%), participants indicated that the pandemic helped them improve information seeking related skills such as “critically reflect upon information sources I find online” (32.2\%), “critically reflect upon information I read online” (31.4\%), “inform myself about current events in Germany and around the world using online news sites” (28.7\%), “purposefully looking for information that is relevant to me” (28\%), “to seek credible sources for online news” (26.6\%), and “to look for appropriate sources to find health-related information (e.g., on Covid-19)” (25.8\%).
\begin{figure*}[h]
\caption{Descriptive means across six age groups concerning improved digital skills during the pandemic}
\centering
\includegraphics[width=0.8\textwidth]{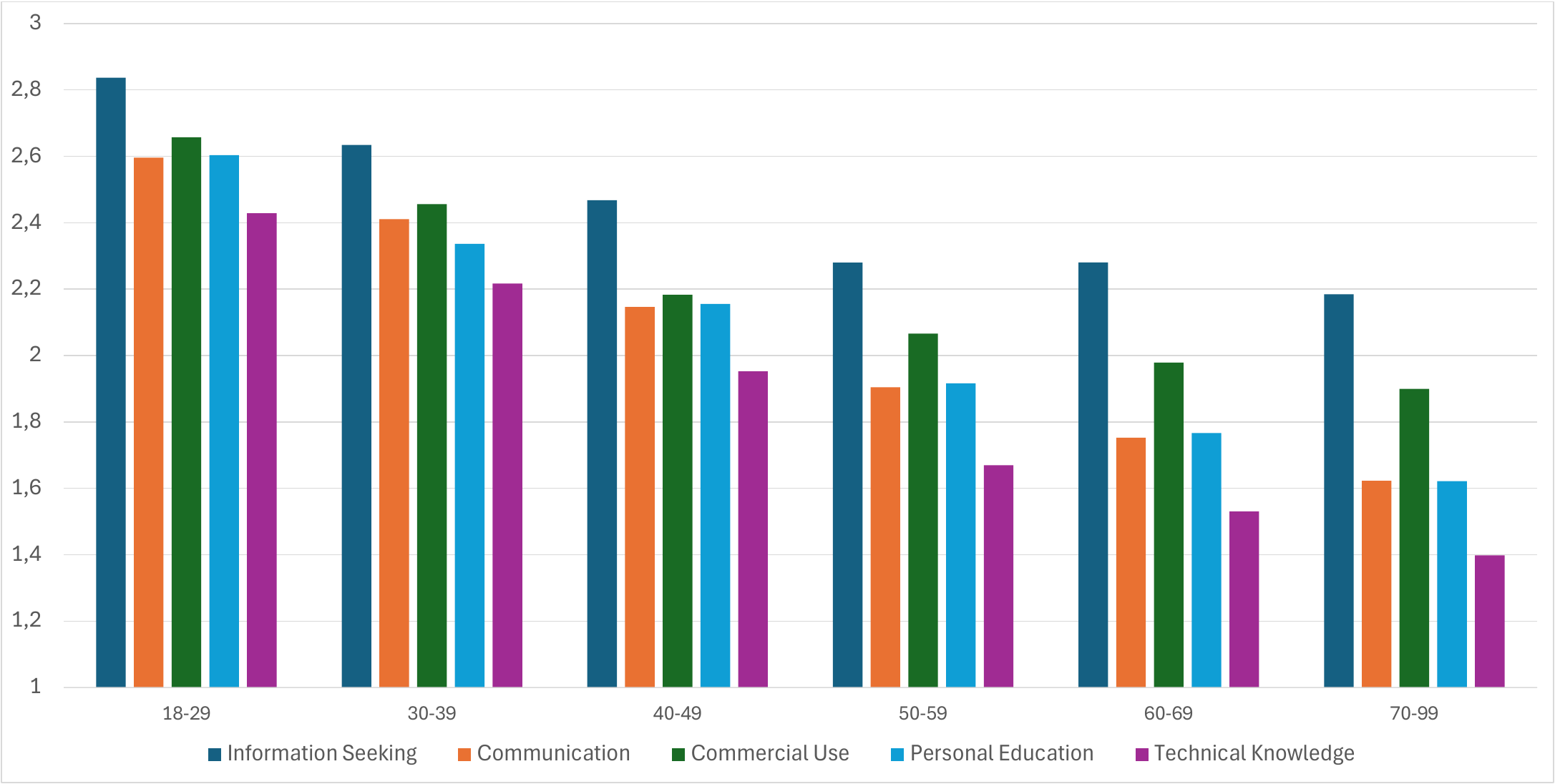}
\label{fig:descriptives1}
\Description{A clustered barplot that displays the distribution of descriptive means of perceived improved digital skills during the pandemic (information seeking, communication, commercial use, personal education, technical knowledge) across six age groups (18-29, 30-39, 40-49, 50-59, 60-69, 70-99)}
\end{figure*}

\subsection{Which socio-demographic factors are associated with improvements in digital skills during the pandemic? (RQ1)}
To operationalize socio-demographic factors, we decided to include age, gender, level of education, household income, and urban verus rural living as these variables are commonly used in the digital divide literature \cite{van_Deursen_2010,elena2021assessing,van_Deursen_2020,hargittai2019internet}.In the interest of maintaining a concise and focused presentation of results, we decided to deviate from our pre-registered plans and do not present the effects of marital status, housing situation, and (German) state of living in this paper.

\textbf{Age.} Correlation analyses revealed that the younger the Internet users, the more likely their digital skills benefited from the pandemic. Most strikingly, age correlated negatively with technical knowledge (r = -.34, CI [-.39; -.29]), communication (r = -.31, CI [-.36; -.26]), personal education (r = -.30, CI [-.35; -.24]), commercial use (r = -.21, CI [-.27; -.15]), and information seeking (r = -.18, CI [-.23; -.12]). Figure \ref{fig:descriptives1} displays that still improvements in informationseeking skills stand out across all age groups.

\textbf{Gender.} A t-test for independent samples indicated that the digital skills of male Internet users benefited from the pandemic more than female Internet users, showing this for technical knowledge, t(1110.48) = –5.98, p $<$ .001, Cohen’s d = –.35 (CI: -.47; -.24), personal education, t(1114.48) = –4.61, p $<$ .001, Cohen’s d = –.27 (CI: -.39; -.16), communication, t(1108.07) = –4.01, p $<$ .001, Cohen’s d = –.24 (CI: -.35; -.12), commercial use, t(1131.43) = –3.93, p $<$ .001, Cohen’s d = –.23 (CI: -.35; -.12), and information seeking, t(1131.76) = –3.51, p $<$ .001, Cohen’s d = –.21 (CI: -.32; -.09). See Table \ref{tab:descriptive_means} for descriptive values.
 
\textbf{Level of Education.} A MANOVA using education (three levels: low, moderate, high) as fixed factor and the five types of digital skills as dependent variables, indicated a multivariate effect, Wilks‘ $\lambda = .97$, F(10,2242) = 3.78, p $<$ .001, $\eta_\rho^2$ = .02, of education. There were further significant univariate effects of education on communication, F(2,1125) = 6.88, p = .001, $\eta_\rho^2$ = .01, technical knowledge, F(2,1125) = 6.81, p = .001, $\eta_\rho^2$ = .01, and personal education, F(2,1125) = 5.52, p = .004, $\eta_\rho^2$ = .01, but not for information seeking, F(2,1125) = .09, p = .913, $\eta_\rho^2$ = .00, or commercial use, F(2,1125) = 1.41, p = .246, $\eta_\rho^2$ = .00. Pairwise comparisons (with Bonferroni correction) revealed that especially Internet users with the highest level of education were more likely to improve their skills in terms of communication, technical knowledge, and personal education than users with low or moderate education (see Table \ref{tab:descriptive_means}) for descriptive means.
 
\textbf{Income.} Computing a MANOVA, we found that income (three levels: low, moderate, high) did not have a multivariate effect on newly improved digital skills, Wilks‘ $\lambda$ = .99, F(10,2260) = 1.76, p = .064, $\eta_\rho^2$ = .01. Still, we found significant univariate effects of income on communication, F(2,1134) = 3.38, p = .035, $\eta_\rho^2$ = .01, personal education, F(2,1134) = 3.75, p = .024, $\eta_\rho^2$ = .01, and technical knowledge, F(2,1134) = 3.10, p = .046, $\eta_\rho^2$ = .01, but not for information seeking, F(2,1134) = 2.20, p = .111, $\eta_\rho^2$ = .00, or commercial use, F(2,1134) = 1.85, p = .158, $\eta_\rho^2$ = .00. Pairwise comparisons (with Bonferroni correction) revealed that especially Internet users with the highest level of income were more likely to improve their technical knowledge compared to those with the lowest level of income (p = .042).
 
\textbf{Urban versus rural living.} A t-test for independent samples indicated that Internet users living in cities were more likely to improve their digital skills in terms of technical knowledge t(793.01) = –5.46, p $<$ .001, Cohen’s d = –.35 (CI: -.47; -.23), personal education, t(832.91) = –5.14, p $<$ .001, Cohen’s d = –.32 (CI: -.44; -.20), communication, t(838.31) = –4.78, p $<$ .001, Cohen’s d = –.30 (CI: -.42; -.18), information seeking, t(910.78) = –4.05, p $<$ .001, Cohen’s d = –.25 (CI: -.37; -.13), and commercial use, t(894.04) = –2.82, p = .002, Cohen’s d = –.17 (CI: -.29; -.06), than those users living in rural areas (e.g., countryside or small town). See Table \ref{tab:descriptive_means} for descriptive values.

\begin{table*}[ht]
\centering

\begin{tabular}{llrlrlrlrlrl}
  \toprule 
& &\multicolumn{2}{c}{\textbf{Information}} &\multicolumn{2}{r}{\textbf{Communication}} & \multicolumn{2}{c}{\textbf{Commercial}}& \multicolumn{2}{c}{\textbf{Personal}} &\multicolumn{2}{c}{\textbf{Technical}}\\
&  &\multicolumn{2}{c}{\textbf{seeking}} &\multicolumn{2}{r}{\textbf{}} & \multicolumn{2}{c}{\textbf{use}}& \multicolumn{2}{c}{\textbf{education}} &\multicolumn{2}{c}{\textbf{knowledge}}\\
 \hline \\[-1.8ex]
\multicolumn{2}{l}{Sex} & M & SD &  M & SD &  M & SD &  M & SD  &M & SD \\
& female&2.34$_a$&1.17&1.99$_a$&0.97&2.10$_a$&1.20&1.97$_a$&1.00&1.73$_a$&0.93 \\
 &male&2.60$_b$&1.27&2.25$_b$&1.17&2.39$_b$&1.30&2.26$_b$&1.17&2.10$_b$&1.11\\
  \hline \\[-1.8ex]
\multicolumn{12}{l}{Level of education} \\
&low&2.48$_a$&1.16&2.05$_a$&1.00&2.24$_a$&1.18&2.03$_a$&1.02&1.83$_a$&0.94\\
&moderate&2.46$_a$&1.20&2.04$_a$&1.03&2.19$_a$&1.23&2.06$_a$&1.04&1.86$_a$&0.99\\ 
&high&2.50$_a$&1.31&2.30$_b$&1.20&2.33$_a$&1.36&2.28$_b$&1.22&2.09$_b$&1.17\\

 \hline \\[-1.8ex]
 \multicolumn{12}{l}{Level of income (monthly household net income)}\\
&below 1,500 EUR&2.47$_a$&1.20&2.04$_a$&1.00&2.17$_a$&1.17&2.01$_a$&1.01&1.78$_a$&0.94\\
&between 1,500 and 2,900 EUR&2.38$_a$&1.19&2.05$_a$&1.06&2.18$_a$&1.21&2.05$_a$&1.03&1.90$_{a,b}$&0.99\\
&higher than 2,900 EUR&2.55$_a$&1.26&2.21$_a$&1.13&2.32$_a$&1.32&2.21$_a$&1.17&1.99$_b$&1.11\\
 \hline \\[-1.8ex]
 \multicolumn{12}{l}{Area of living}\\

&rural&2.36$_a$&1.18&2.00$_a$&0.99&2.16$_a$&1.20&1.98$_a$&0.99&1.78$_a$&0.91 \\
&urban&2.66$_b$&1.28&2.32$_b$&1.19&2.38$_b$&1.33&2.33$_b$&1.21&2.14$_b$&1.18 \\

 \hline
   
\end{tabular}
\caption{Descriptive means across five dimensions of improved digital skills during the pandemic. Means having different subscripts (a vs. b) indicate statistically significant differences. When comparing multiple means, we ran Bonferroni-corrected post hoc comparisons with a stricter significance testing to reduce the risk that we falsely identify significant differences due to chance (i.e., Type I Error)}.
\label{tab:descriptive_means}
\Description{The table presents the descriptive means of improved digital skills. The columns represent the improved digital skills (namely, information seeking, communication, commercial use, personal education and technical knowledge) and the rows represent the participants' attributes, namely, sex (female, male), level of education (low, moderate, high), level of income (below 1500 euros, between 1500 and 2900 euros, and higher than 2900 euros), and the area of living (rural, urban).}
\end{table*}

\subsection{How did newly improved digital skills exacerbate or attenuate the psychological effects of the pandemic? (RQ2)}
 
When examining the diversity of different (psychological) effects Internet users attributed to their use of digital media, it becomes clear that feeling informed (M = 3.46, SD = 1.07; 52.4\% agreed/fully agreed), connected to others (M = 3.20, SD = 1.17; 42.8\% agreed/fully agreed), educated (M = 3.05, SD = 1.10; 33\% agreed/fully agreed), and competent (M = 3.03, SD = 1.06; 31.1\% agreed/fully agreed) were among the most intensely perceived effects. We conducted an exploratory factor analysis (principal axis analysis with varimax rotation) whose empirical eigenvalues were used for a subsequent parallel analysis that indicated a two-factor solution. We then calculated a second exploratory factor analysis with an oblique rotation (principal axis analysis with promax rotation) and the fixed number of two factors. This two-factor solution represented two sets of effects due to the use of digital media in individuals’ views: positive (e.g., feeling safe, informed, educated, competent) and negative (e.g., feeling isolated, rejected, under pressure, excluded). Correlation analysis (see Table \ref{tab:correlations}) indicated that while newly improved digital skills helped expand all positive effects, they also contributed to the negative ones. Still, the relationship with positive effects was greater in magnitude than those with negative effects.

\begin{table}[ht]
\centering

\begin{tabular}{lll}
  \toprule 

&Positive effects&Negative effects\\ \hline
Information seeking&.401**&.235**\\
Communication&.449**&.261**\\
Commercial use&.373**&.255**\\
Personal education&.470**&.247**\\
Technical knowledge&.457**&.272**\\

 \hline
   
\end{tabular}
\caption{Correlation coefficients between improved digital skills and positive/negative effects digital media use during the pandemic ($^{**}p<0.01$)}
\label{tab:correlations}
\Description{Table 2 shows the correlation coefficients between the improved digital skills and positive or negative effects of digital media use during the pandemic. The columns signify the positive and negative effects and the rows stand for the improved digital skills (namely, information seeking, communication, commercial use, personal education and technical knowledge) }
\end{table}
 
\subsection{Which are the digital skills of tomorrow from the public’s point of view? (RQ3)}

To provide an overview of what Internet users think will be the most important digital skills of tomorrow, we collapsed the two highest points of the scale ("needed", and "very much needed"). According to participants’ views, the digital skills that will be needed most urgently in the future are: to differentiate between misinformation and accurate information (86.1\%), to distinguish serious from unserious sources on the Internet (85.9\%), to estimate the consequences of one's actions on the Internet (84.8\%), to protect personal data online (84.1\%), and to use social media in a reflective manner (81.1\%) (Table \ref{tab:futuredigitalskills}). 

\begin{table*}[ht]
\centering

\begin{tabular}{lr}
  \toprule 
 Tomorrow's digital skills & Percentage (\%)\\
 \hline \\[-1.8ex]
To differentiate between misinformation and accurate information &	86,1\\
To distinguish serious from unserious sources on the Internet &	85,9\\
To estimate the consequences of one's actions on the Internet&	84,8\\
To protect personal data online	&84,1\\
To use social media in a reflective manner&	81,1\\
To estimate when and how online information can lead to wrong conclusions&	80,3\\
To purposefully look for information&	79,6\\
To judge one's own abilities using digital media&	79,1\\
To thoughtfully share/spread information&	79\\
To protect other people's data online&	77,7\\
To virtually communicate with others in a polite and an appropriate way&	76\\
To consume accurate and balanced information on the Internet&	73,4\\
To handle artificial intelligence such as ChatGPT&	60,9\\
To use virtual rooms constructively&	50,5\\
To interact with algorithms&	46,1\\
To create virtual rooms in line with one's preferences&	38,7\\
 \hline
   
\end{tabular}
\caption{Percentages of two highest points of the scale ("needed", "very much needed") referring to tomorrow’s digital skills.}
\label{tab:futuredigitalskills}
\Description{Table 3 shows the percentages of participants' responses regarding the most important digital skills of tomorrow}
\end{table*}

\subsection{Who - in Internet users’ view - is responsible for fostering these digital skills? (RQ4)}
 
\begin{figure*}[h]
\caption{Descriptive overview of who is/should be responsible for fostering individuals’ digital skills.}
\centering
\includegraphics[width=0.8\textwidth]{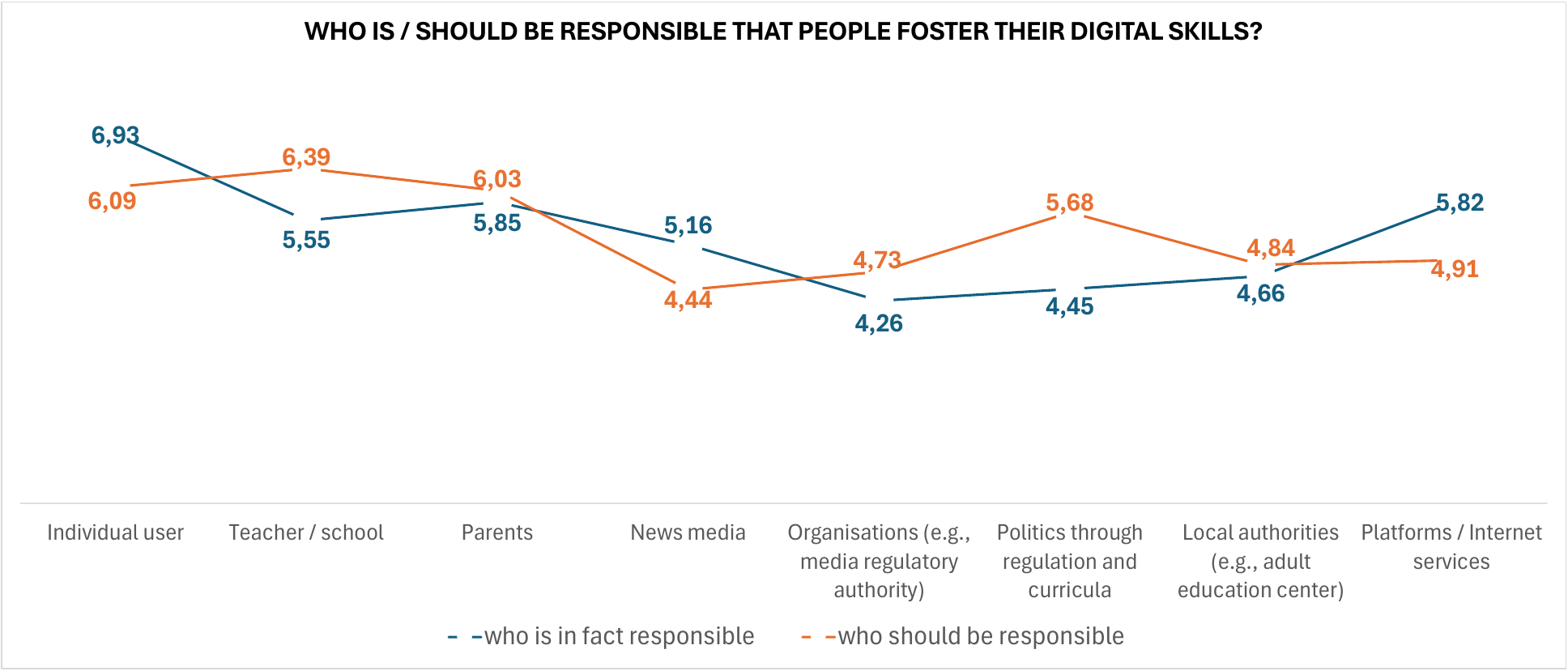}
\label{fig:fosteringskills}
\Description{The figure presents a comparison of participants' responses regarding who is responsible for fostering digital skills and who should be responsible.}
\end{figure*}

When it comes to placing the responsibility for fostering digital skills, participants saw the individual user, schools/teachers, and parents as the ones most in charge (see Figure \ref{fig:fosteringskills}; note that higher means represent the attribution of more responsibility). Comparing who is currently responsible with who should be responsible, according to the participants' perceptions, it is evident that participants would like to shift the responsibility from individual users to formalized spaces of education such as schools. Participants also saw political officials having a prominent role in regulating and providing education plans so that people could effectively acquire digital skills.

\section{Discussion}
This research intended to examine a) the distribution of newly improved digital skills due to COVID-19 lockdowns among the population, b) the association of those skills with psychological effects of COVID-19 that were extended by digital media, and c) users' view on the digital skills the accelerated digital transformation demands from individuals. Our results suggest that COVID-19 did not help to close gaps in the digital divide. Instead, the COVID-19-related accelerated digital transformation of daily life helped advantaged groups to expand their digital skills further. Nevertheless, we did not find any indications suggesting that digital transformation due to the pandemic was harmful to the less digitally skilled users. 

\subsection{COVID-19 and the second-level digital divide: A few beneficiaries, but no disadvantaged}
In line with previous assessments of the second-level digital divide \cite{Correa_2015,Mart_nez_Cantos_2023,Oeldorf_Hirsch_2023,_zsoy_2020,van_Deursen_2010}, we found that young, male, and higher educated Internet users' subjectively benefited the most from the pandemic by improving their digital skills compared to female, older, and lower educated individuals. These findings could be attributed to the fact that the groups who expressed stronger improvements in their digital skills were the groups found to have used the Internet in a more intense form and sometimes even in new ways during the pandemic \cite{Nguyen_2021b,van_Deursen_2020,mcclain2021internet}. Given that the latter results stemmed from countries other than Germany indicates that our present findings might be generalizable to other populations. The universality of these results becomes apparent when exemplarily examining the effects of area of living: The notion that people living in urban areas subjectively improved their digital skills during the pandemic (more than those living in rural areas; as shown in the German sample) could be because urban residents estimated the Internet as more essential in their lives during the pandemic than rural residents (as shown in a U.S. American sample \cite{mcclain2021internet}). The rural-urban digital divide, though, might be attributable to other socio-economic factors (such as level of education) than just geography \cite{hindman2000rural}. Still, we see the pattern that enhanced use of the Internet and estimates of technology as key during the pandemic go hand in hand with improved digital skills. 
When focusing on the different dimensions of digital skills, it is an interesting observation that while gender made a difference for all dimensions, the level of education most strikingly determined improvements in "communication," "personal education," and "technical knowledge." This might be due to how differently people with higher educational levels have used digital media during the pandemic compared to those with lower educational levels. For instance, PEW research indicated that adults with advanced degrees were more likely to have daily video calls and to use technology in new forms than adults with lower educational degrees \cite{mcclain2021internet}.

Against the expected potential outcome, the accelerated digital transformation of social, educational, and work-life did not close any gaps in the digital divide. Still, it did not harm the disadvantaged groups either. In all hitherto disadvantaged groups (female, older, and lower-educated participants), we observed an improvement in digital skills, meaning that we cannot assess an exacerbation of the disadvantaged. The fact that those who had already sophisticated skills before the pandemic had a greater chance to improve their digital skills during the pandemic could still result in a widening gap in the digital divide concerning digital skills. A key question is how those groups managed their digital skills after the pandemic. Were those who newly improved their digital skills (e.g., female and older Internet users) due to the pandemic able to maintain those skills and use them in their daily life, or do we observe an extension of the digital divide because one group improved while the other returned to the original status?

\subsection{Digital skills and their role in buffering psychological effects of COVID-19}
As research documented that (digital) media played a beneficiary role in people's coping with the pandemic \cite{Eden_2020,Nabi_2022,Wulf_2022,Drazich_2022,Koban_2022,Sun_2023}, it is pivotal to assess that all forms of digital skills helped to expand those positive effects. It seems that knowing how to evaluate information, online sources, how to connect with others, or how to technically handle daily life purchases contributed to people feeling more safe, satisfied, and competent during the pandemic. This underscores that it is not simply the use of digital media that helps people cope with extreme situations, but rather their ability to use these media effectively to harness the positive effects. An enhanced ability to use technology effectively, in turn, makes individuals self-determined actors in a highly connected world, as their technology use allows them to fulfill basic psychological needs—such as the need for autonomy—which they may subjectively perceive as a gratifying experience \cite{dietrich2024constitutes, kermavnar2024assessing}.
For instance,  perceiving autonomy through technology and exerting control over their communication seems especially important and psychologically functional in times, like the COVID-19 pandemic, when individuals have little control over their external circumstances. On the other hand, considering that the use of technology was also documented to have a negative impact (e.g., daytime tiredness, anxiety \cite{Drazich_2022,Koban_2022,Sun_2023}), digital skills were found to go hand in hand with the negative effects. The most obvious negative psychological effect due to enhanced technology use is represented by "Zoom fatigue" and the notion that people - while expanding their digital skills when making video calls - also felt worn out and stressed by the newly emerging intensity of technology-mediated communication \cite{mcclain2021internet,deniz2022zoom}. Moreover, knowing how to handle information (e.g., on the spread of the virus) online, thus, could lead individuals to develop stronger anxiety and more realistic views about the severity of this virus as well as, ultimately, the long-term consequences for individuals and whole societies. Similarly, it seems conceivable that being distracted through efficiently used digital media helped for a while, but when returning to the pandemic reality, individuals face issues with re-adapting to this situation. Therefore, as digital media usage itself, having digital skills appears to have both benefits and drawbacks for handling extreme situations, such as global health crises. According to the statistical correlations, the beneficial effects of digital skills, though, were larger than the detrimental ones.

\subsection{The nature of newly improved digital skills and those needed for the future}
While connecting with family members, friends, or co-workers through video conference software -- such as Zoom -- seems to be one of the key activities people engaged with digital media \cite{Karl_2021,Juvonen_2021,Nguyen_2021,Scott_2022}, it is remarkable that developing skills to better judge information and sources in digital media was the second most prevalent digital skill provoked due to the pandemic and mentioned as an important set of skills for future Internet users. It seems that the pandemic made the infodemic, that is, the increasing spread of misinformation, very salient to users \cite{gerosa2021mis,Gabarron_2021,Freiling_2021,Rocha_2021}, making them also aware of the importance of this skill. As shown by our present results, the enhanced use of the Internet due to the pandemic made individuals improve their information-seeking skills. While males made greater improvements than females, individuals' self-perceived amelioration of information-seeking skills was independent of the level of education.  In our study, participants expressed that individual users are responsible for developing these skills but still should be relieved by formalized interventions provided by educational institutions and regulated by political officials. While some of the skill dimensions are easier to improve through self-directed or peer-supported learning (e.g., communication through videoconferencing), others require more formalized forms of learning (e.g., technical knowledge), holding official institutions accountable to support the further distribution of these skills and knowledge. 

\subsection{Theoretical and Practical Implications}

The findings presented in this paper have a number of implications for theory and practice. Our findings provide insights into the impact of global challenges, such as health crises, on people's skills and capabilities relating to technology and, potentially, psychological consequences and societal implications.

Global health crises, such as the COVID-19 pandemic, despite the major negative impact on individuals and society, can also act as accelerators for individuals to acquire or practice specific skills. Policymakers, state officials, and educators could use these insights to design guidelines for preparing and recovering from global crises to minimize and balance the negative impact on the population. Furthermore, these insights offer suggestions for regulating the digital divide, for example, by identifying population groups with specific characteristics and needs and offering accessible and sustainable life-long training opportunities.

Additionally, our results suggest that the proficiency and experience of people with digital technologies can have an impact on the psychological consequences that people faced during the pandemic. This can inform and guide healthcare stakeholders who issue guidelines for dealing with the psychological effects of global crises.

Our study provides additional information regarding the digital skills that users themselves consider important for navigating digital spaces and the roles of official organizations and institutions in promoting digital literacy. This information is especially critical nowadays with the ongoing discussions regarding the role and impact of emerging technologies, such as AI and generative AI, and the need for regulation in specific application areas, such as education \cite{holmes2022artificial} and overall \cite{Leslie_2021}. We agree that regulation is needed to safeguard against the potential negative impacts of emerging technologies. However, we align with voices that argue that promoting education and fostering the acquisition of the necessary skills to use and engage meaningfully with technology is even more important and necessary to ensure harnessing the potential of technology and solving the difficult challenges ahead \cite{gredel2024education}.  

\subsection{Limitations}
First, a pivotal limitation of this survey is that we measured participants' self-perceived and not the actual improvement of digital skills. We intended to close the gap between subjective and objective levels of skills by presenting participants with very specific questions about their activities and associated abilities. In fact, previous research showed that self-assessments of digital skills significantly correlate with actual digital skills and that this relationship is high when assessment refers to operational digital skills \cite{deursen2010measuring}. Still, there is a difference between both levels of digital skills that needs to be addressed by future research, also assessing the evolution of digital skills in the aftermath of the pandemic. Second, while our interpretation of results refers to a potential widening gap of the second-level digital divide, the actual extension of the gap could only be observable over time by assessing whether improvements become larger for one group over time while they do not for the other. Based on the framing of our items, we intended to grasp this trend through the measurement. Still, further research needs to follow up on this potential trend. Third, this survey did not cover a potential decline in digital skills as we asked participants only for subjective improvements. While the enhanced technology use due to the pandemic may not directly suggest declines in people's digital skills, it seems conceivable that the forced nature of technology use made individuals realize their limited abilities. Fourth, further socio-demographic factors such as the particular occupation (not surveyed in the present study) could serve as key determinants of the extent of improvements in digital skills. For instance, those whose work was fully digitized before the pandemic may not have improved their digital skills compared to those whose working environment turned digital due to the pandemic. These considerations should be examined in future research.

\section{Conclusion}
This study extends our understanding of the distribution of digital skills and, more strikingly, how the acquisition of digital skills can be accelerated by extreme circumstances such as a global pandemic inducing a digital transformation. While the more intense use of technology during the pandemic did not close any gaps in the distribution of digital skills, it generally helped to improve Internet users' digital skills (especially their communication and information-seeking skills). For those who were already advantaged, this improvement manifested itself to a stronger extent. Especially the ability to evaluate information and sources thoroughly stands out as a newly improved skill that users also evaluate as an important skill for members of digital societies to develop. 
\begin{acks}
The authors would like to thank the Interdisziplinäres Zentrum für Bildungsforschung (IZfB) at the University of Duisburg-Essen for funding and supporting this research.
\end{acks}

\bibliographystyle{ACM-Reference-Format}
\bibliography{references}




\end{document}